\documentclass[aip, jmp, amsmath,amssymb,preprint]{revtex4-1}

\usepackage{graphicx}
\usepackage{dcolumn}
\usepackage{bm}
\begin{document}
\title{Wavebreaking amplitudes in warm, inhomogeneous plasmas revisited} 

\author{Nidhi Rathee}
\email[]{nidhi.rathee@ipr.res.in}
\affiliation{Institute for Plasma Research, Bhat, Gandhinagar, Gujarat, 382428, India}
\affiliation{Homi Bhabha National Institute, Training School Complex, Mumbai, 400094, India}

\author{Arghya Mukherjee}
\altaffiliation[Present affiliation: ]{Center of Excellence in Space Sciences India, Indian Institute of Science Education and Research Kolkata, Mohanpur-741246, India}
\affiliation{Institute for Plasma Research, Bhat, Gandhinagar, Gujarat, 382428, India}
\affiliation{Homi Bhabha National Institute, Training School Complex, Mumbai, 400094, India}

\author{R.M.G.M. Trines}
\affiliation{Central Laser Facility, STFC Rutherford Appleton Laboratory, Didcot, OX11 0QX, United Kingdom}

\author{Sudip Sengupta}
\affiliation{Institute for Plasma Research, Bhat, Gandhinagar, Gujarat, 382428, India}
\affiliation{Homi Bhabha National Institute, Training School Complex, Mumbai, 400094, India}

\date{\today}

\begin{abstract}
The effect of electron temperature on the space-time evolution of nonlinear plasma oscillations in an inhomogeneous plasma is studied using a one-dimensional particle-in-cell (PIC) code. In contrast to the conventional wisdom, it is found that for an inhomogeneous plasma, there exists a critical value of electron temperature beyond which wave breaking does not occur. This novel result, which is of relevance to present day laser plasma interaction experiments, has been explained on the basis of interplay between electron thermal pressure and background inhomogeneity.
\end{abstract}


\maketitle 

\section{Introduction}
The determination of maximum sustainable amplitude by nonlinear oscillation/waves in a plasma has been a topic of fundamental interest since last six decades\cite{akhiezer56,Davidson72,kruer88,gibbon05}. Apart from academic interest, the breaking of nonlinear plasma waves is also a topic of practical importance to a number of systems \textit{viz.} particle acceleration experiments\cite{tajima79,modena95,malka02,hegelich06,schwoerer06,faure06,matlis06}, laser assisted fusion schemes\cite{tabak94,kodama01}, collisionless heating of laboratory plasma \cite{koch74,bose15,bauer92,sandhu05}, heating of solar corona \cite{botha00,voitenko05,hasegawa74} etc. Dawson\cite{dawson59} first theoretically demonstrated that the maximum electric field amplitude that can be sustained by nonlinear electron oscillations in a cold homogeneous plasma, is given by $kA=1$ where $A = eE_{max}/m\omega_{0}^2$ is the amplitude of electron's oscillation,  $\omega_{0}$ is electron plasma frequency with homogeneous background density $n_0$, $m$ is mass of electron and $k$ is the wave number. At this amplitude electron density becomes singular at a particular point in space and the wave breaks within a period. As a consequence, the coherent wave energy gets transformed to the random particle energy which heats up the plasma. 

Dawson\cite{dawson59} further extended his calculations by introducing an inhomogeneity in the ambient plasma density and showed that inclusion of time independent plasma inhomogeneity (due to ions) completely changes the dynamics of the excited oscillations/waves. 
It was found that, inclusion of background inhomogeneity inevitably results in breaking of the plasma oscillations at arbitrarily low amplitudes. 
Physically, in an inhomogeneous plasma the characteristic plasma frequency of oscillation becomes space dependent ( \textit{i.e} $\omega_{pe}(x) = [4\pi n_i(x)e^2/m]^{1/2}$ where $n_i(x)$ is ambient inhomogeneous plasma density), due to which neighboring oscillating electrons gradually go out of phase and eventually cross each other resulting in breaking of the coherent motion. The electron density exhibits a singularity at this point. The energy which is initially loaded on a long wavelength mode gradually goes into higher and higher  harmonics\cite{kaw73}, and eventually the condition of wave breaking
$k_{wb}A = 1$\cite{dawson59} is satisfied. This phenomenon is known as phase mixing\cite{drake76,infeld_relativistic,sudip99prl,sandhu05,
sudip09pre,sudip11ppcf,prabalprl,chandanprl,arghya14} which results in wave breaking ( wave breaking via phase mixing ); and the time the oscillations take to break is known as phase mixing time ( wave breaking time ), which depends on the strength of inhomogeneity as has been shown by several authors\cite{infeld89,infeld90,nappi91,mithun18,mithunscripta}.  

All the above studies were carried out for a cold plasma. In 1971, Coffey\cite{coffey71} analytically studied the breaking of nonlinear electron plasma waves in a warm homogeneous plasma. Using a water bag model\cite{Davidson72} for electrons, Coffey showed that for a warm plasma, the wave breaking amplitude explicitly depends on the electron temperature as $eE_{max}/m\omega_{pe}v_{\phi}=(1-8\beta^{1/4}/3+2\beta^{1/2}-\beta/3)
^{1/2}$ where $\beta = 3v_{th}^2/v_{\phi}^2$  is the normalized electron temperature and $v_\phi$ is the phase velocity. 
It was clear from Coffey's calculation that inclusion of finite electron temperature significantly reduces the wave breaking amplitude and prevents density singularity due to the thermal pressure\cite{coffey71,moriscripta,bauer92,kruer_scripta,kruer79}.
In case of driven plasma waves in a spatially inhomogeneous plasma also, finite electron temperature reduces the wave breaking amplitude, as found by Kruer \textit{et. al.}\cite{kruer79}.
Later studies, for self-excited plasma waves, by Infeld \textit{et. al.}\cite{infeld89},
on the combined effect of plasma inhomogeneity and electron temperature on the space time evolution of nonlinear electron plasma waves led to the conclusion  that electron density acquires a maximum value $n_{max}/n_0 = 1/(1-\alpha^{3/2}\gamma^{-1/2}/k_i\lambda_{D})$ in time $T_{b} = 2 \pi K(1/\sqrt(2))/k_{i}\lambda_{D}(\gamma \alpha)^{1/2}$, which Infeld defined as wave breaking time. Here $K$ is the complete elliptic integral, $k_{i}$ and $\alpha$ are respectively the wave number and amplitude of the ion inhomogeneity, $\gamma = c_{p}/c_{v}$ and
$\lambda_{D}$ is the electron Debye length. It is clear from the  expression for density maximum that its validity is limited to $k_{i}\lambda_{D} > \alpha^{3/2} / \gamma^{1/2}$; it was not clear from reference \cite{infeld89} as to what would happen, if this condition is violated. 

In 2009, Trines\cite{trines09} revisited this problem and revealed the interplay between the thermal pressure and phase mixing due to inhomogeneity. It was found that in an inhomogeneous plasma, thermal pressure not only reduces the wave breaking amplitude, but beyond a critical value may entirely prevent the onset of wave breaking. Physically it happens as follows: density inhomogeneity on its own causes growth and ``accumulation" of $k's$ at a given spatial location (referred to as secular effect in reference \cite{trines09}) resulting in peaking of density while thermal effects cause advection of the $k's$ resulting in smoothening of density profile. If the rate of advection of $k's$ is greater than the rate at which the $k's$ grow and ``accumulate", wave breaking will not occur and maximum $k$ gets limited to a value ($k_{max}$) which is determined by the electron temperature and the inhomogeneity amplitude.

To elucidate the aforementioned interplay between growth in $k's$  due to inhomogeneity and advection of $k's$ due to thermal effects, in this paper we  present 1-D PIC simulations\cite{birdsall85} of nonlinear electron plasma oscillations in the presence of inhomogeneous stationary ions. We find that the fate of the excited oscillation is essentially decided by the maximum wave number ($k_{max}$) generated in the system, which in turn is governed by the electron temperature and the inhomogeneity amplitude.
We observe that, if the maximum value ($k_{max}$) is less than $k_{wb}$ \textit{i.e.} $k_{max} < k_{wb}$ ( where $k_{wb} A \approx 1$; where $A$ is the amplitude of the wave excited by the background inhomogeneity), wave breaking will not occur. The $k's$ essentially ``walk-off'' the density gradient resulting in smoothening of the density profile. Wave breaking occurs only if the condition $k_{max} > k_{wb}$ is satisfied. 

This paper is organized as follows: in section \ref{theory}, we present a theoretical analysis on the spatio-temporal evolution of electron wave number ($k$), and discuss it's connection with the previously known results by Infeld {\it et. al.}\cite{infeld89} and Trines\cite{trines09}. Further we present the condition at which wave-breaking can be avoided. Later in section \ref{simulation} we present our simulation results for the same. Finally in section \ref{summary}, we conclude and summarize our results.

\section{Theoretical analysis}\label{theory}

In this section, we present theoretical 
 analysis on the effect of electron temperature on electron plasma oscillations in an inhomogeneous plasma. We start from a plasma with background electron density $n_0$, an immobile sinusoidal ion density inhomogeneity of the form $n_i(x,t) = n_0[1 + \alpha \cos (k_ix)]$, Debye length $\lambda_{D} = v_{th}/\omega_{0}$, and $\gamma = c_p/c_v > 1$. Presence of ion inhomogeneity results in the excitation of an electron plasma 
oscillation with amplitude $A = \alpha / k_{i}$ and initial wave number $k_i$. Spatio-temporal evolution of this excited wave is governed by the Eikonal equation\cite{whitham74,landau72}, which can be expressed as $\partial k/\partial t+\partial\omega/\partial x=0$, where the term 
$\partial\omega/\partial x$ is evaluated using the warm plasma dispersion relation. For an inhomogeneous warm plasma, the Bohm-Gross\cite{bohm49} dispersion relation may be approximately written as $\omega^2 \approx \omega_{pe}^2(x)+\gamma k^2v_{th}^2$, where $\omega$ is the frequency of oscillation, $k$ is the wave number, $\omega_{pe}^2(x)=4\pi e^2 n_{i}(x)/m_e$ is the plasma frequency (which is space dependent in our study), $\gamma = c_p/c_v$, and $v_{th}=\omega_{0}\lambda_{D}$ is the thermal velocity of electrons ( $\omega_{0}$ is plasma frequency corresponding to $n_0$ ). 
Substituting the expression for $\omega$ from the dispersion relation,  the Eikonal equation transforms to 

\begin{equation}
\frac{\partial k}{\partial t}+\frac{\gamma v_{th}^2k}{\omega_{0}}\frac{\partial k}{\partial x} \approx -\frac{\omega_{0}}{2 n_0}\frac{\partial n_i}{\partial x}   \label{equation1}
\end{equation}
Following Trines\cite{trines09} we transform the coordinates as $t = \tau $ and $x = x' +(\gamma v_{th}^2/\omega_{0})\int_0^{\tau}{kd\tau'}$, which leads to 
\begin{equation}
\frac{\partial k}{\partial \tau} \approx -\frac{\omega_{0}}{2 n_0}\frac{\partial n_i}{\partial x}   \label{equation2}
\end{equation}
Integrating Eq.\ref{equation2}, we finally get
\begin{equation}
 \triangle (k\lambda_{D})^2\leqslant \triangle n_i / (\gamma n_0)  \label{equation3}
\end{equation}
which gives an upper bound on the wave number $k$; for a sinusoidal inhomogeneity, as chosen above, the upper bound on wave number $k$ turns out to be 
\begin{equation}
k_{max} \approx \sqrt{\alpha/\gamma}/\lambda_{D}   \label{equation4}
\end{equation}
As stated in the introduction, the above value of $k_{max}$ is determined by the phenomenon that $k$ advects with the group speed, and will eventually ``walk off'' the density gradient. Since the growth of $k$ is caused by that same density gradient, this ``walk-off'' will curb the growth of $k$. Associated with the spatio-temporal evolution of wave number $k$, the density maximum of the excited Langmuir wave will not only grow because of growth in $k$, but also smoothen because of advection of $k$ (via the group speed $\gamma v_{th}^2 k / \omega_{0}$). This means that the growth will saturate once the maximum $k$ ( $k_{max}$ ) has ``walked off'' the density gradient. The ``walk off'' time which we denote by $T_{w}$, is the time $k$ takes to reach the upper bound $k_{max}$, and can be estimated from Eq.\ref{equation2} itself, which for a sinusoidal inhomogeneity turns out to be 
\begin{equation}
\omega_{0}T_{w}\approx 2/k_i\lambda_{D}\sqrt{\alpha\gamma} \label{equation5}
\end{equation}
Also the density maximum at the walk-off time $T_{w}$ can be estimated using $n_{max} \approx n_{0}/(1 - k_{max}A)$, which finally leads to (using $A = \alpha / k_{i}$ )
 \begin{equation}
\frac{n_{max}}{n_0} \approx \frac{1}{1-\frac{\alpha^{3/2}\gamma^{-1/2}}{k_i\lambda_{D}}} 
\label{equation6}
 \end{equation}
We note here that the expression for $T_{w}$ and $n_{max}$ are similar to the unnumbered expression for $T_{b}$ and $n_{e,max}$ at the end of reference \cite{infeld89}. We further note that Eq.\ref{equation6}, can only be used if saturation of $k$ due to Eq.\ref{equation4} happens before wave breaking due to $k_{wb} A \approx 1$ occurs.  
This is because, the upper bound on the wave number, $k_{max}\lambda_{D} \approx \sqrt{\alpha/\gamma}$, is found from the maximum possible growth of $k$ on a slope of the ripple in the ion density, and not from any limitation imposed by wave breaking. Thus we need to separately determine whether the wave will break within the time $T_{w}$, or just reach a finite maximum density (given by Eq. \ref{equation6}) and stay unbroken after that. 

Following Dawson \citep{dawson59}, we define wave breaking time $T_{b}$ as $T_{b} = k_{wb}/(\partial k / \partial t)$. For a cold plasma with sinusoidal inhomogeneity (where ``walk-off'' time $T_{w} \rightarrow \infty$ ), using $k_{wb} = 1/ A = k_{i} / \alpha$ and $\partial k / \partial t \approx \omega_{0} k_{i} \alpha / 2 $, the wave breaking time $T_{b}$ turns out to be $\omega_{0} T_{b} \sim 2/\alpha^{2}$ ( a finite value). Thus we recover the well known result that nonlinear electron oscillations in a cold inhomogeneous plasma will always break as $T_{b} < T_{w}$. In terms of wave number this is $k_{wb} < k_{max} $ (for a cold plasma $k_{max} \rightarrow \infty$ {\it i.e.}  the wave breaking limit is reached before ''walk-off'' is complete ). 
We note here that wave breaking in a warm inhomogeneous plasma may be avoided by simply changing the sign of the aforementioned inequality {\it i.e} $T_{b} > T_{w} $ ( or $k_{wb} > k_{max} $ ). Using the approximate condition $k_{wb} A \approx 1$ for a warm plasma, this inequality again translates into the condition $k_{i}\lambda_{D} > \alpha^{3/2}/\gamma^{1/2}$.  This is exactly the same condition which makes the density maximum given by Eq. \ref{equation6} finite and well defined, indicating a well defined, unbroken plasma wave. Under these condtions the maximum value that $k$ can reach (obtained from Eq.\ref{equation4}) is less then $k_{wb}$ i.e. $k_{max} < k_{wb}$, implying that $k$ will never reach $k_{wb}$ and wave will simply ``walk-off'' (never break) and the peak electron density acquires a value given by the expression found by Infeld {\it et. al.}\cite{infeld89} ( Eq. \ref{equation6} ).

We now address the issue of peak plasma density at wave breaking in a warm inhomogeneous plasma. Use of the approximate condition $k_{wb} A = 1$, for the warm plasma case leads to a density singularity as $n_{max} \approx n_0 / ( 1 - k_{wb} A )$. Unlike the cold plasma case where density goes to infinity as wave breaks, physically in a warm plasma density cannot become singular. This is because a divergent plasma density will lead to a divergent plasma pressure which in turn will not allow the electron density to become singular. As argued by Trines\cite{trines09}, inclusion of full nonlinear plasma pressure leads to the following modified wave breaking condition in a warm plasma
\begin{equation}
k_{wb}A \approx 1-(\gamma k_{wb}^2 \lambda_{D}^2)^{1/(\gamma +1)} \label{equation7}
\end{equation}
Using the above wave breaking condition, maximum electron density at wave breaking may be written as 
 \begin{equation}
 \frac{n_{max}}{n_0} \approx \frac{1}{1-k_{wb}A}=\frac{1}{(\gamma k_{wb}^2 \lambda_{D}^2)^{1/(\gamma + 1)}} 
 \label{equation8}
 \end{equation}
which is well defined and finite.
Finally the wave breaking time for the warm inhomogeneous case may be estimated from Eq.\ref{equation2} by integrating it within the limit $k=0$ to $k=k_{wb}$, and can be written as 
\begin{equation}
\omega_{0}T_{b}\approx 2k_{wb}/\alpha k_i \label{equation9}
\end{equation}
As discussed above wave breaking will occur if $T_{b} < T_{w}$. For a  $1-D$ plasma, using $\gamma=3$, Eq.\ref{equation7} can be solved for $k_{wb}$ as 
\[ k_{wb}=\frac{k_i}{\alpha}\left[ 1 - \sqrt{3}k_i\lambda_{D}
\left\lbrace (1 + 4\alpha / \sqrt{3}k_i\lambda_{D})^{1/2}-1\right\rbrace/2\alpha \right] \]
which when substituted in Eq.\ref{equation9} yields the expression for wave breaking time $T_{b}$, as  
\begin{equation}
 \omega_{0}T_{b}\approx \frac{2}{\alpha^2}\left[ 1 - \sqrt{3}k_i\lambda_{D}
\left\lbrace (1 + 4\alpha/\sqrt{3}k_i\lambda_{D})^{1/2}-1\right\rbrace/2\alpha \right] \label{equation10}
 \end{equation}

 We would like to point out here that there is a subtle difference between the approaches to wave breaking as described in Trines\cite{trines09} and that described above. But both the approaches are entirely consistent with each other. In Trines\cite{trines09}, it is assumed that $k_{max}$ is given (it is determined by the amplitude ``$\alpha$'' of the ion inhomogeneity and the electron temperature) and then proceeds to determine the value that the amplitude $A$ can assume before the wave breaks. $A$ takes the wave breaking amplitude $A_{wb}$ when the wave breaking condition in a warm plasma {\it i.e.} $k_{max}A_{wb} = 1 - (\gamma k_{max}^{2} \lambda_{D}^{2})^{1/(\gamma + 1)}$ is satisfied. This gives Eq.(5) of Trines\cite{trines09}. In our description, the initial amplitude $A$ is given ( it is determined by the amplitude ``$\alpha$'' and the wave number ``$k_{i}$'' of the ion inhomogeneity), and we then determine the value for $k$ required to trigger wave breaking, and if or when this value is reached. $k$ takes the value $k_{wb}$ when the wave breaking  condition in a warm plasma {\it i.e.} $k_{wb}A = 1 - (\gamma k_{wb}^{2} \lambda_{D}^{2})^{1/(\gamma + 1)}$ (Eq.\ref{equation7} above) is satisfied. Instead of evaluating $k_{wb}$, by putting $k_{wb} = k_{max}$ in the above equation and using this to determine $A$ ({\it i.e.} $A_{wb}$) leads to Eq. (5) of Trines\cite{trines09}. Thus the theory of wave breaking as described in our work is consistent with Trines\cite{trines09}. 

\section{Simulation results}\label{simulation}

In this section we present our simulation results obtained using an $1-D$ Particle in Cell (PIC) code. We perform PIC simulations of nonlinear electron plasma oscillations in an inhomogeneous warm plasma and compare the simulation results with the theoretical predictions presented in the previous section. In simulation we have used the following normalization: $t\rightarrow\omega_{0}t$, $x\rightarrow k_i x$, and $v_e\rightarrow k_i v_e/\omega_{0}$, $n_e\rightarrow n_e/n_0$, $E\rightarrow k_i eE/m\omega_{0}^2$.

 The parameters for our simulations are as follows: The system length is chosen to be 
$L = 2 \pi$, which implies $k_{i} = 1$. It is divided into $NG$ equal sized cells, where the cell size is chosen to be of order $\Delta x = L/NG \approx \lambda_{D}$ in order to avoid non-physical instabilities\cite{kruerpic,birdsall85}.
The number of particles per cell is taken as $\sim 100$ and the time step for numerical integration is taken as $\sim 5 \times 10^{-3}$ ($\omega_{0} \Delta t \ll 1 $).
Electrons are loaded with a Maxwellian distribution with electron temperature ($v_{th}$) using the inversion technique described in reference \cite{birdsall85}. 
  
Starting from initial conditions provided in the previous section, we follow the spatio-temporal evolution of electron plasma oscillations ( of amplitude $A = \alpha/k_{i}$) for several hundreds of plasma period. It is observed that as time progresses the electron density begins to peak and reaches a maximum value which explicitly depends on the inhomogeneity $\alpha$ and electron temperature $v_{th}$.
We have carried out two sets of simulations. First set of simulations is carried out with constant $v_{th} = 0.01$ ( Figs. 1 - 5 ), which implies that $k_i \lambda_D = 0.01$ ( $NG = 512$ so that $\Delta x \sim \lambda_{D}$ ). The initial amplitude of the self-excited plasma wave is $A = \alpha/k_i$. To study the effect of inhomogeneity on wave-breaking, $\alpha$ is varied from $0.01 - 0.08$ ($A$ also varies in the same range), which implies that the parameter $\alpha^{3/2}/\gamma^{1/2}$ ranges from $5.8 \times 10^{-4} - 0.013$. Therefore as $\alpha$ varies from $0.01 - 0.08$, the inequality $k_{i} \lambda_{D} > \alpha^{3/2}/\gamma^{1/2}$ ( $k_{max} < k_{wb}$ ) switches sign and as per theory described in section \ref{theory}, we expect to move from no wave-breaking regime to a wave breaking regime. 
In the other set of simulations (Figs. 6 - 10), we have kept $\alpha = 0.05$ constant, which implies that $\alpha^{3/2}/\gamma^{1/2} \approx 6.5 \times 10^{-3}$, and $A = \alpha/k_i = 0.05$. To study the effect of temperature on wave-breaking $v_{th}$ is varied from $0.001 - 0.05$, which implies that 
$k_{i} \lambda_{D}$ varies from $0.001 - 0.05$. ( In order to avoid non-physical instabilities, the number of cell $NG$ is varied accordingly,
such that $\Delta x \sim \lambda_D$ for each value of $v_{th}$ ). Therefore as $v_{th}$ varies from $0.001 - 0.05$, the inequality $k_{i} \lambda_{D} < \alpha^{3/2}/\gamma^{1/2}$ ( $k_{max} > k_{wb}$ ) switches sign and we expect to move from wave-breaking regime to a no wave breaking regime.

\begin{figure}%
\includegraphics[width=0.9\linewidth,height=0.5\linewidth]{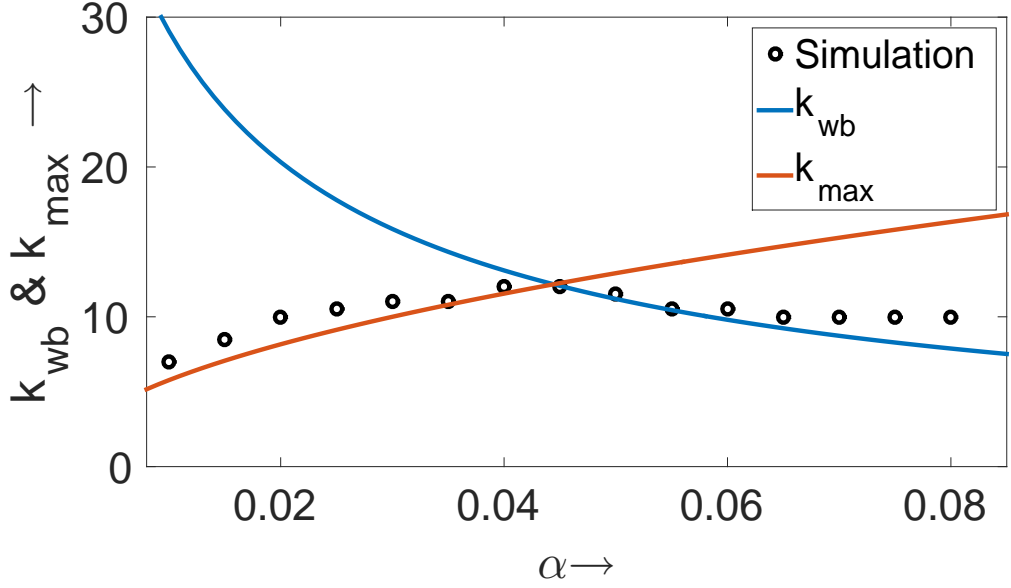}
\caption{Comparison of $k_{wb}$ and $k_{max}$ as function of $\alpha$ for $v_{th}=0.01$. Dots represent the maximum value of $k$ obtained from simulation.}
\label{fig:1}
\end{figure}

\begin{figure}%
\includegraphics[width=0.9\linewidth,height=0.5\linewidth]{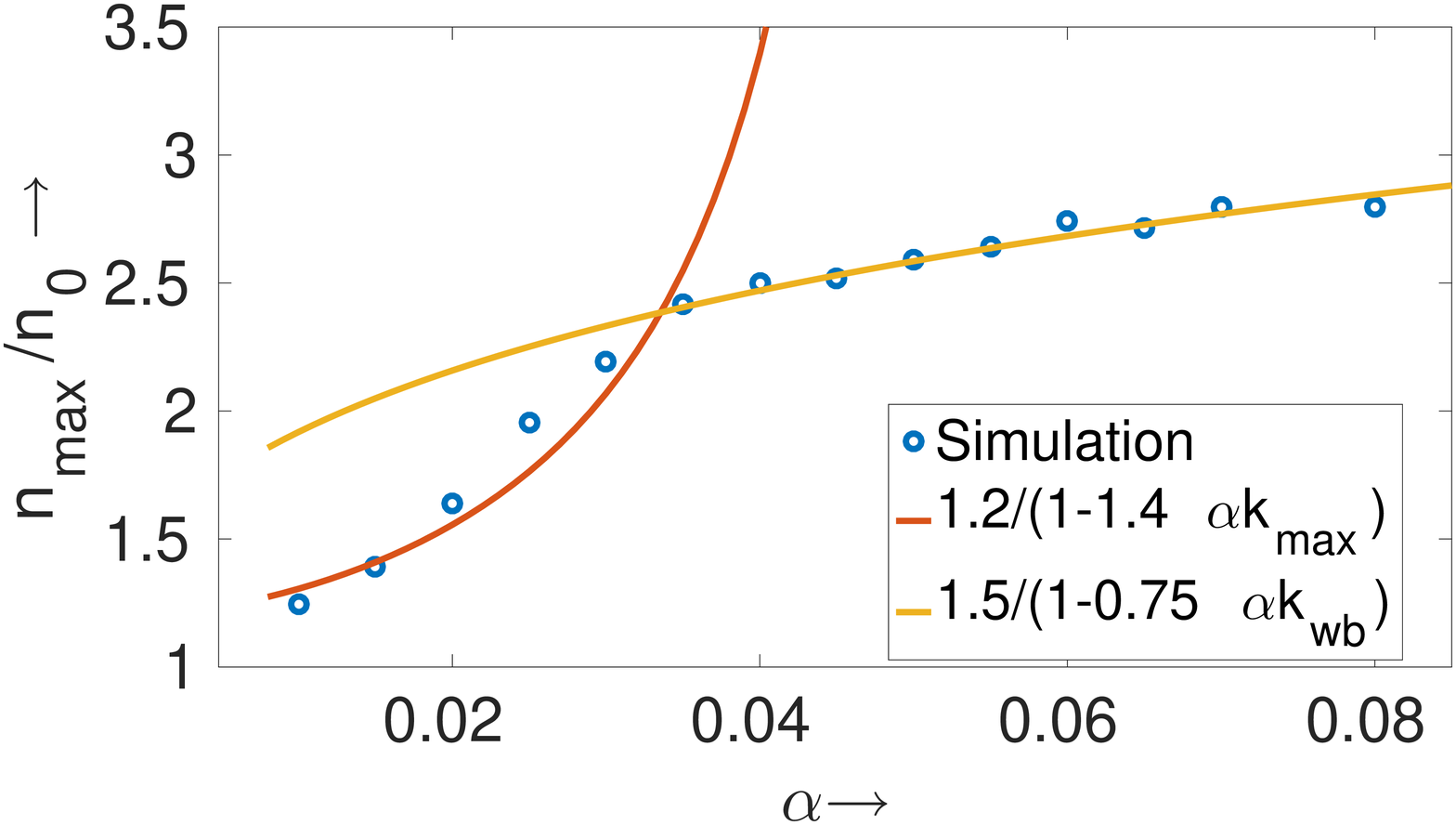}
\caption{Maximum normalized density as a function of $\alpha$ for $v_{th}=0.01$. The dots represent the simulation results and the lines represent the theoretical scalings discussed in text.}
\label{fig:2}
\end{figure}

 We now present our simulation results. We first compare the theoretically obtained expressions for $k_{max}$ (Eq.\ref{equation4}) and $k_{wb}$ (Eq.\ref{equation7}) with simulation results. Fig.\ref{fig:1} shows the variation of $k_{wb}$ and $k_{max}$ as function of ion inhomogeneity $\alpha$ for a constant $v_{th}=0.01$ (here $k_i\lambda_{D}$ is constant). In this figure the theoretical value of $k_{wb}$ and $k_{max}$ are respectively shown by blue and orange line. Dots represent the value of maximum wave number obtained from simulation. 
We have recorded the mode number which has the highest amplitude at the time when the electron density is maximum, and taken that as maximum $k$ value. Our results show that for $\alpha \lesssim 0.0445$, {\it i.e.} when the theoretical value of $k_{max} <  k_{wb}$, the maximum value of $k$ obtained from simulation closely follows the theoretical $k_{max}$ curve, whereas when the value of $\alpha$ is increased beyond $0.0445$, the maximum simulation $k$ closely follows the theoretical $k_{wb}$ curve. The electron density maximum also shows the expected scaling with $\alpha$ (see fig.\ref{fig:2}),  albeit the form of density scaling is found to change at a little lower value of $\alpha$ (lower than $0.0445$).
\begin{figure}%
\includegraphics[width=0.9\linewidth,height=0.5\linewidth]{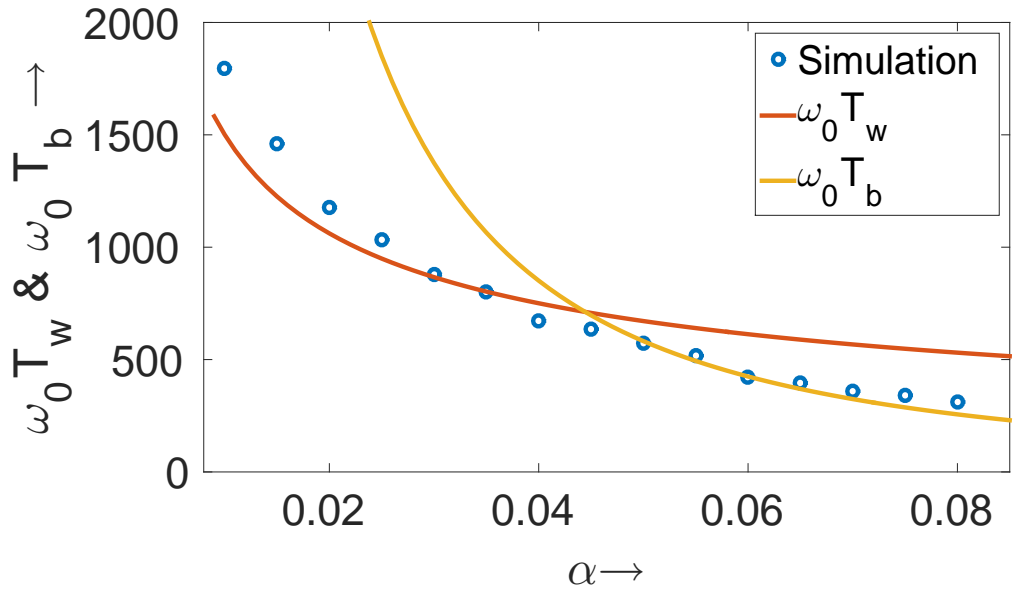}
\caption{The time at which maximum density is achieved as a function of $\alpha$ for $v_{th}=0.01$. The dots represent the simulation results and the lines represent the theoretical scalings discussed in text.}
\label{fig:3}
\end{figure}%
In fig.\ref{fig:2}, we show the variation of maximum electron density $n_{max}/n_0$ as a function of $\alpha$, for a fixed electron temperature ($v_{th} = 0.01$). 
We find that if $\alpha \lesssim 0.035$, 
the maximum electron density (shown by dots) obtained from simulation follows the scaling given by Eq.\ref{equation6},
as shown by orange curve. Fig.\ref{fig:2} also shows that the density maximum (shown by dots) follows a different scaling given by Eq.\ref{equation8} (shown by yellow curve) for $\alpha \gtrsim 0.035$, 
 We note here that while fitting the density expressions with the simulation results, the scalings given by Eq.\ref{equation6} and Eq.\ref{equation8} have been used in a phenomenological fashion as $n_{max}/n_{0} \approx c_{1}  / (1 - c_2 k A)$, where $c_{1}$ and $c_{2}$ are fitting constants of order unity ($c_1 \approx 1.2$ and $c_2 \approx 1.4$, for $k = k_{max}$ and $c_1 \approx 1.5$ and $c_2 \approx 0.75$ for $k = k_{wb}$). 
Fig.\ref{fig:3} shows the variation of the time taken to reach the density maximum as a function of inhomogeneity amplitude $\alpha$, for a fixed value of $v_{th} = 0.01$. Here also, the dots represent the simulation results and the solid lines are theoretical predictions given by Eq.\ref{equation5} (orange curve for $k_{max} < k_{wb}$) and Eq.\ref{equation10} (yellow curve for $k_{max} > k_{wb}$). Here, to fit the simulation results, Eq.\ref{equation5} and Eq.\ref{equation10} have been multiplied with a constant of order unity $\sim 1.3$.
 
\begin{figure}%
\includegraphics[width=1.0\linewidth,height=0.5\linewidth]{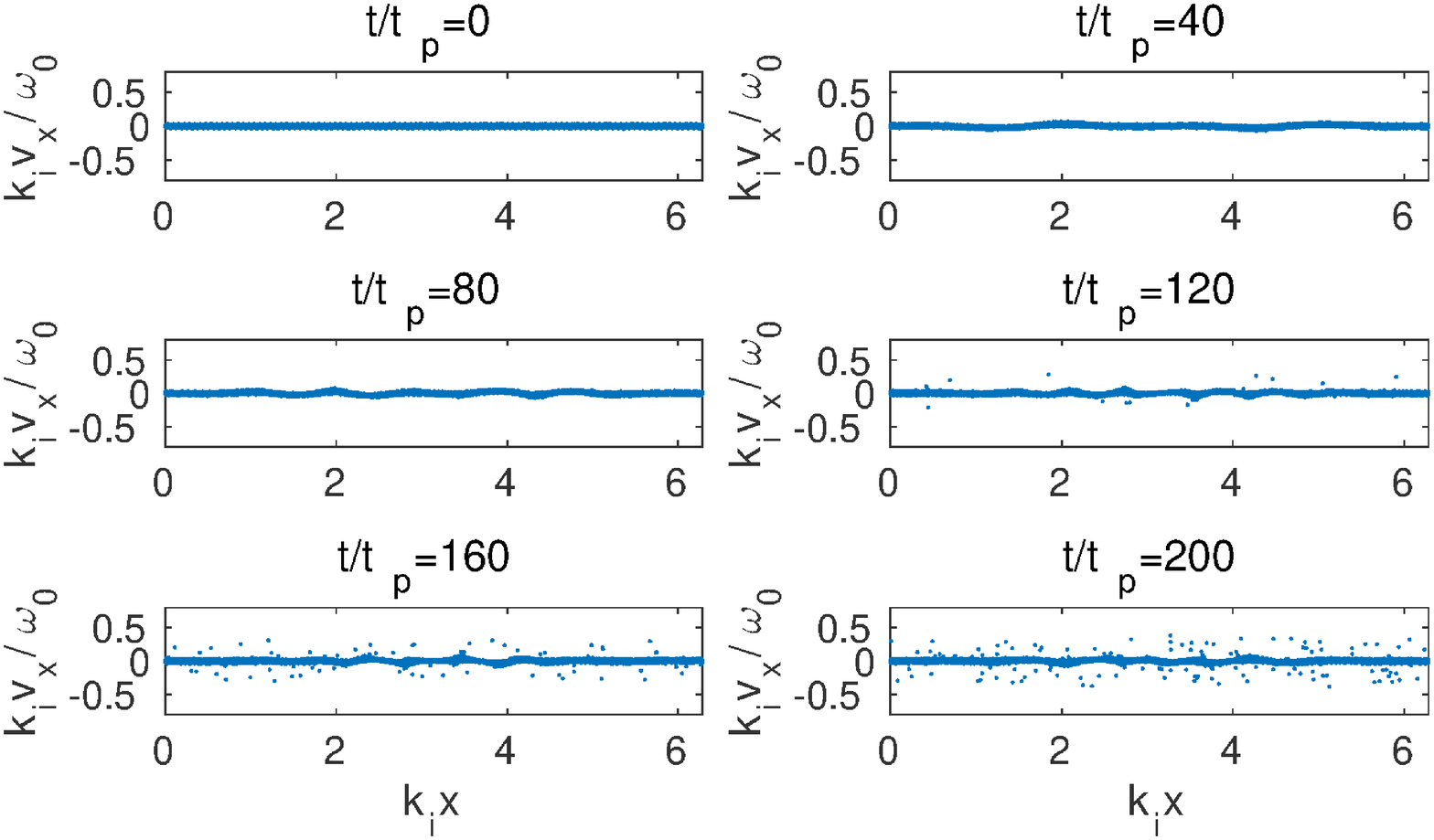}
\caption{Phase space plot for $v_{th}=0.01$ and $\alpha=0.03$ ( $ < 0.0445$, $k_{max} < k_{wb}$ ).}
\label{fig:4}
\includegraphics[width=1.0\linewidth,height=0.5\linewidth]{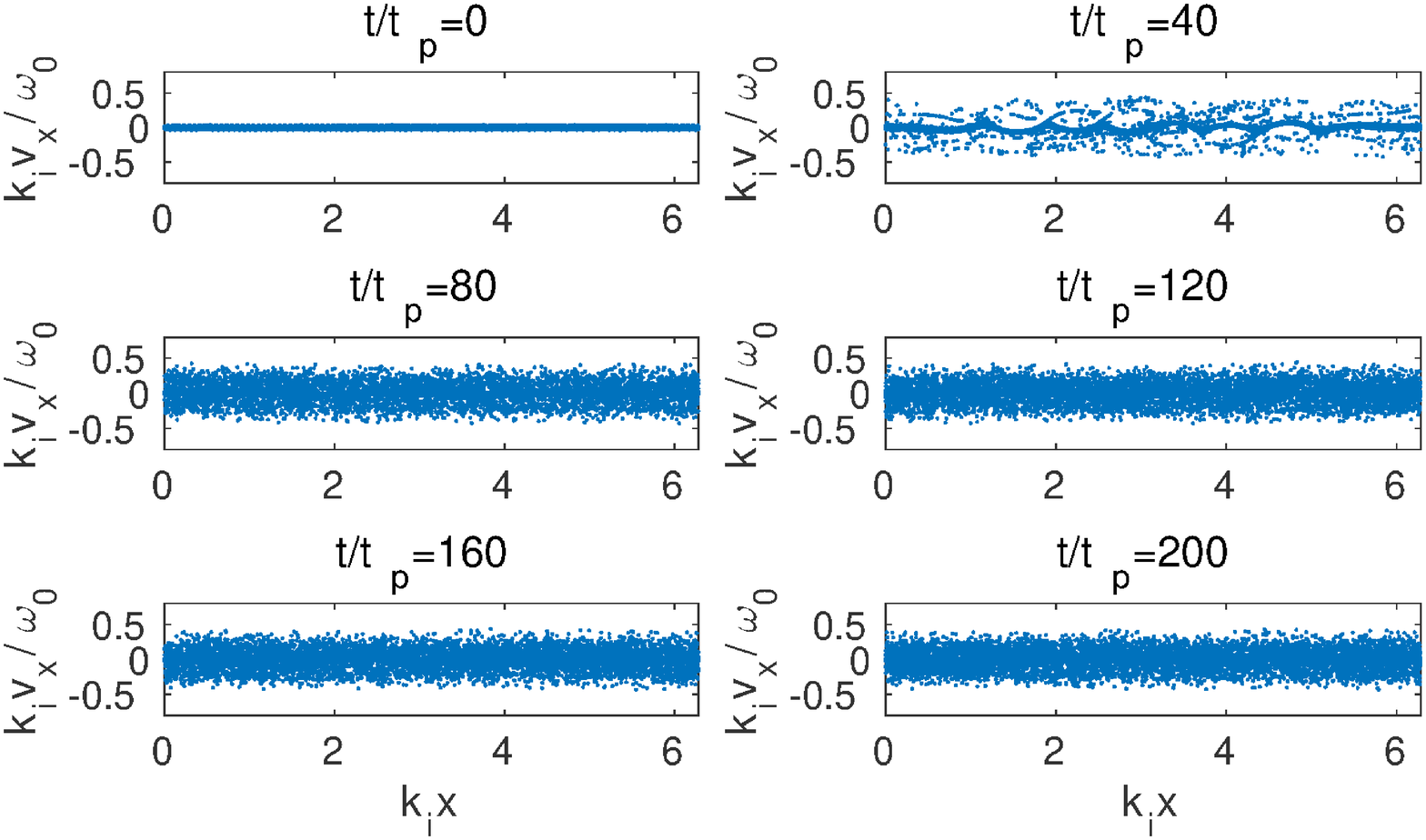}
\caption{Phase space plot for $v_{th}=0.01$ and $\alpha=0.08$ ( $ > 0.0445$, $k_{max} > k_{wb}$ ).}
\label{fig:5}
\end{figure}

In order to distinguish between the two regimes {\it i.e} when wave breaking does not occur ($k_{max} < k_{wb}$) and when wave breaking does occur ($k_{max} > k_{wb}$), we present snapshots of the electron phase space for two different values of $\alpha$ for a fixed value of thermal velocity $v_{th} = 0.01$. Fig.\ref{fig:4} and \ref{fig:5} respectively show snapshots of the electron phase space for same thermal velocity $v_{th}=0.01$, for two different inhomogeneity amplitudes $\alpha=0.03$ ( $ < 0.0445$, $k_{max} < k_{wb}$ ) $\&$ $0.08$ ( $ > 0.0445$, $k_{max} > k_{wb}$ ), at different times. At $\omega_{0}t=0$, the velocity spread is same in both cases. After few plasma periods it is observed that, for $\alpha=0.08$, the width of the phase space increases significantly. It implies that a large fraction of the wave energy is transferred to particles in the form of random kinetic energy. We identify this as wave breaking, as large number of particles are getting accelerated to high velocities. Whereas for $\alpha=0.03$ the phase space does not change much from it's initial shape and the number of energetic particles are much smaller in number as compared to that seen in Fig.\ref{fig:5}. We conclude that for $\alpha=0.08$ the wave breaks where as for $\alpha=0.03$ the wave does not break (even at longer times the phase space doesn't change), as expected from the theoretical analysis presented in the previous section.

\begin{figure}%
\includegraphics[width=0.9\linewidth,height=0.5\linewidth]{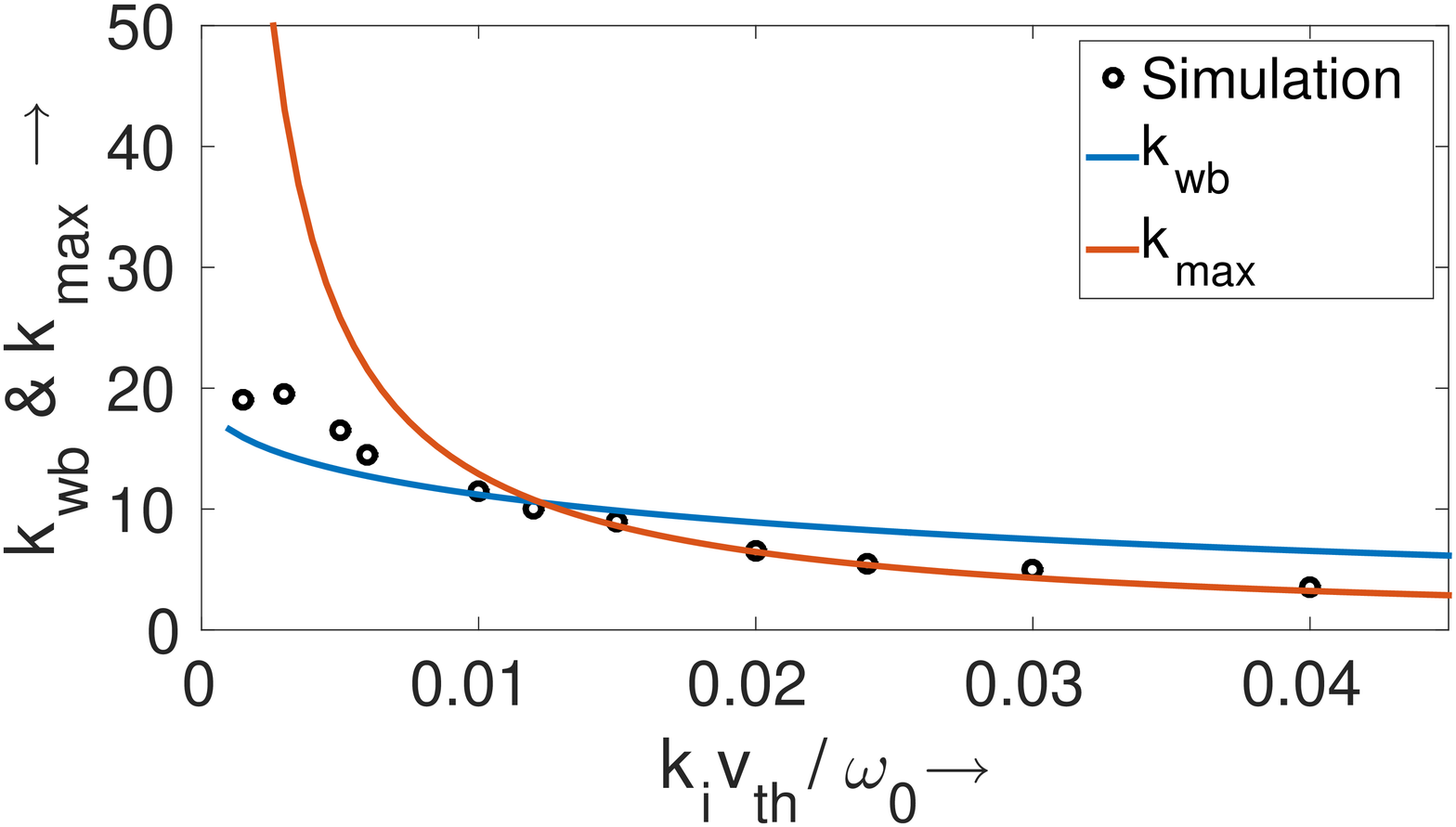}
\caption{Comparison of $k_{wb}$ and $k_{max}$ as a function of $v_{th}$ for $\alpha=0.05$. Dots represent the maximum value of $k$ obtained from simulation.}
\label{fig:6}
\end{figure}

\begin{figure}%
\includegraphics[width=0.9\linewidth,height=0.5\linewidth]{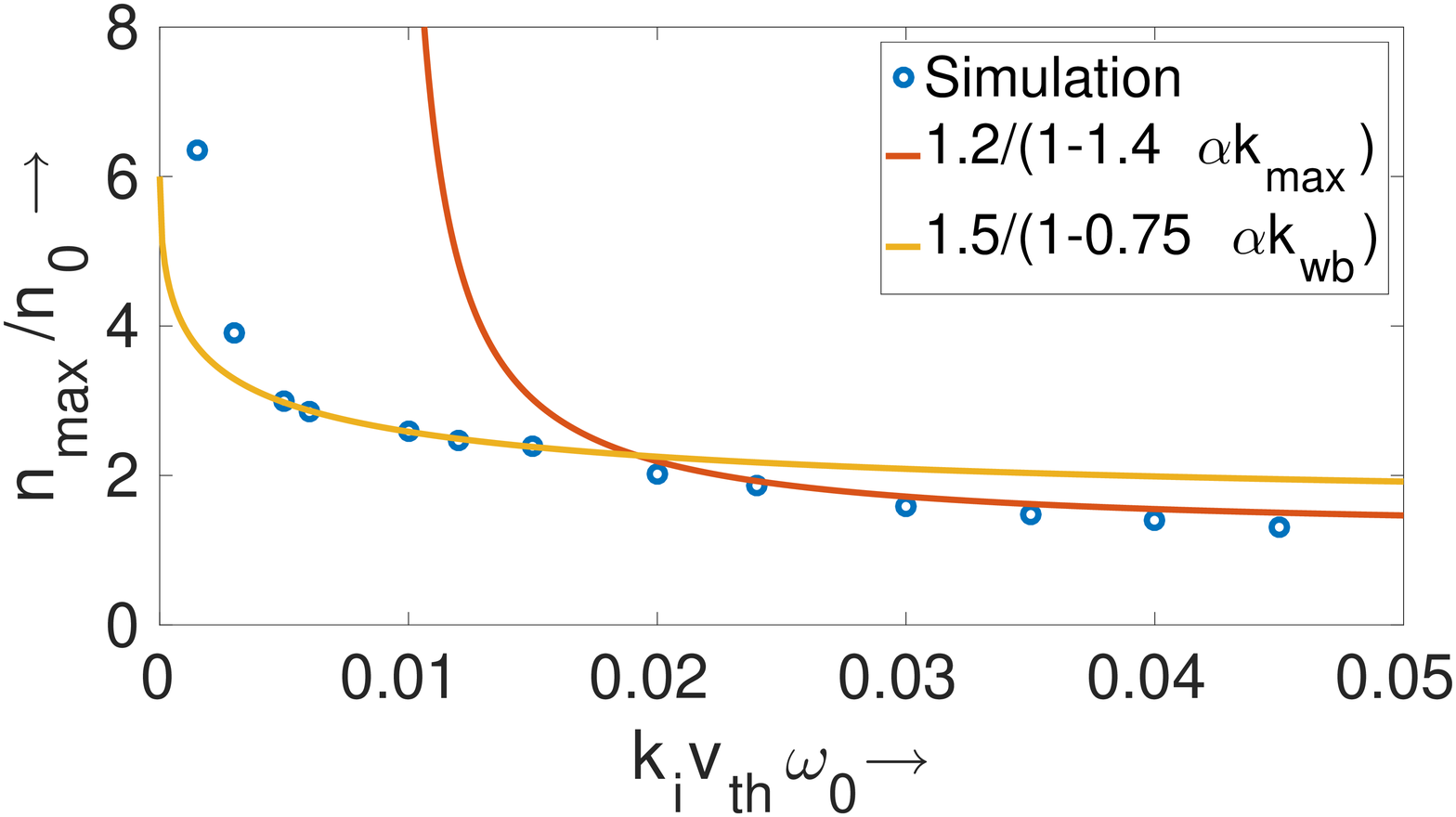}
\caption{Maximum normalized density as a function of $v_{th}$ for $\alpha=0.05$. The dots represent the simulation results and the lines represent the theoretical scalings discussed in text.}
\label{fig:7}
\end{figure}

\begin{figure}%
\includegraphics[width=0.9\linewidth,height=0.5\linewidth]{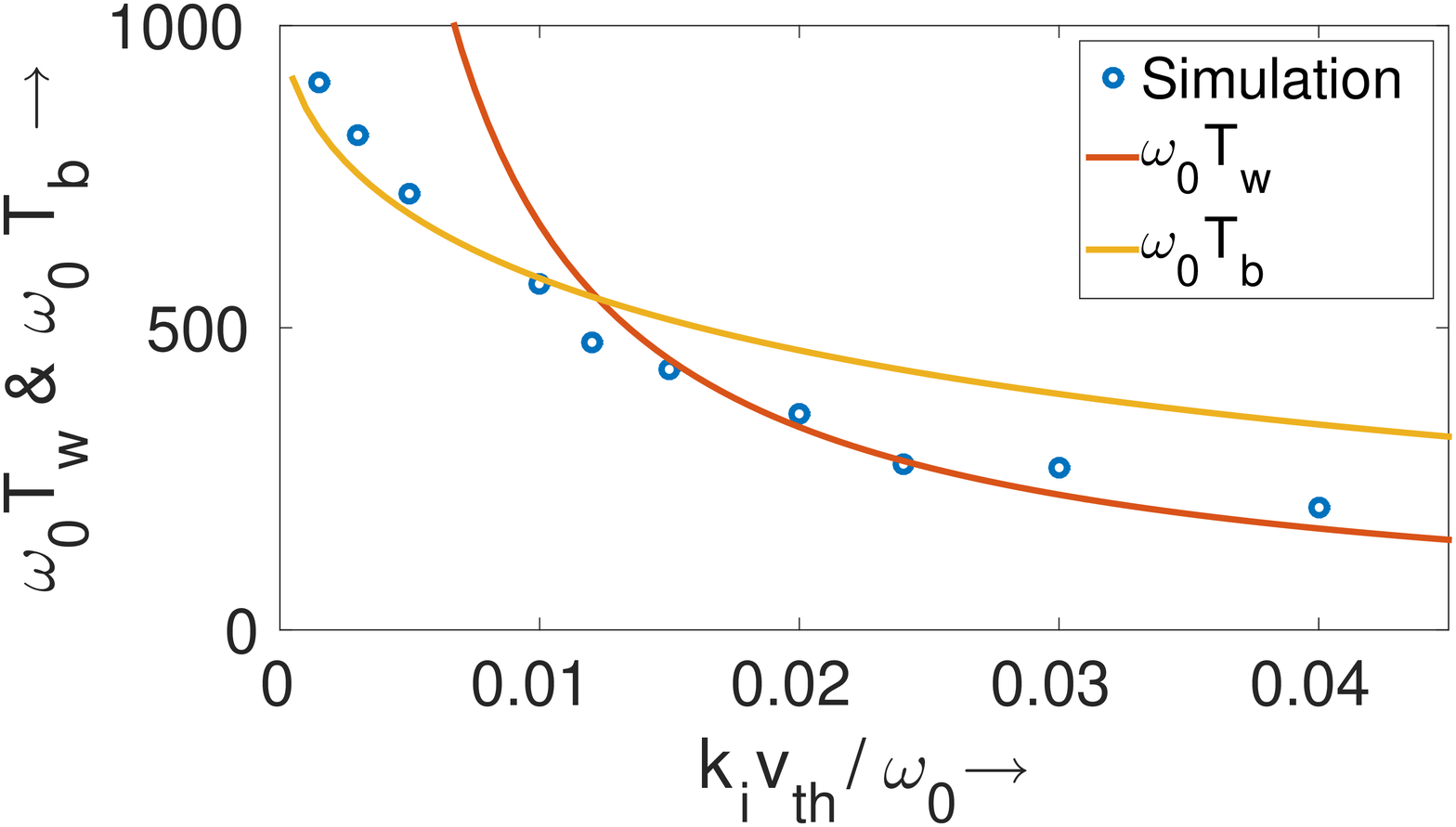}
\caption{Time taken to reach the maximum density as a function of $v_{th}$ for $\alpha=0.05$. The dots represent the simulation results and the lines represent the theoretical scalings discussed in text.}
\label{fig:8}
\end{figure}

Next we keep the inhomogeneity parameter $\alpha$ fixed and vary $v_{th}$  (for this case $k_i\lambda_{D}$ changes as $v_{th}$ changes). Fig.\ref{fig:6} shows the variation of $k_{wb}$ and $k_{max}$ as a function of $v_{th}$ for a constant value of $\alpha=0.05$. Here also theoretical values of $k_{wb}$ and $k_{max}$ are shown by blue and orange line respectively; while dots represent the maximum value of $k$ obtained from simulation at wave breaking time ($k_{max} > k_{wb}$) or at walk-off time ($k_{max} < k_{wb}$).  
%
In fig.\ref{fig:7} we show the variation of density maximum (with dots) as a function of $v_{th}$ for $\alpha=0.05$. 
 Like the previous case, we again find that density maximum follows two scalings, {\it i.e.} for
higher values of $v_{th}$, the density maximum follows the scaling given by Eq.\ref{equation6} (shown by orange line), whereas for lower values of $v_{th}$, the density maximum follows the scaling given by Eq.\ref{equation8} (shown by yellow line). Our simulation results show that the value of $v_{th}$ at which the density scaling changes is a little higher than the value at which the wave number scaling changes (see Fig.\ref{fig:6}). 
In fig.\ref{fig:8} we present the variation of the time taken to reach this maximum density as function of $v_{th}$. Here also, the dots represent the simulation results and the solid lines are theoretical predictions given by Eq.\ref{equation5} (orange curve for $k_{max} < k_{wb}$) and Eq.\ref{equation10} (yellow curve for $k_{max} > k_{wb}$).  We note that the fitting constants used in Fig.\ref{fig:7} and Fig.\ref{fig:8} are the same as used in Fig.\ref{fig:2} and Fig.\ref{fig:3} respectively.


\begin{figure}%
\includegraphics[width=1.0\linewidth,height=0.5\linewidth]{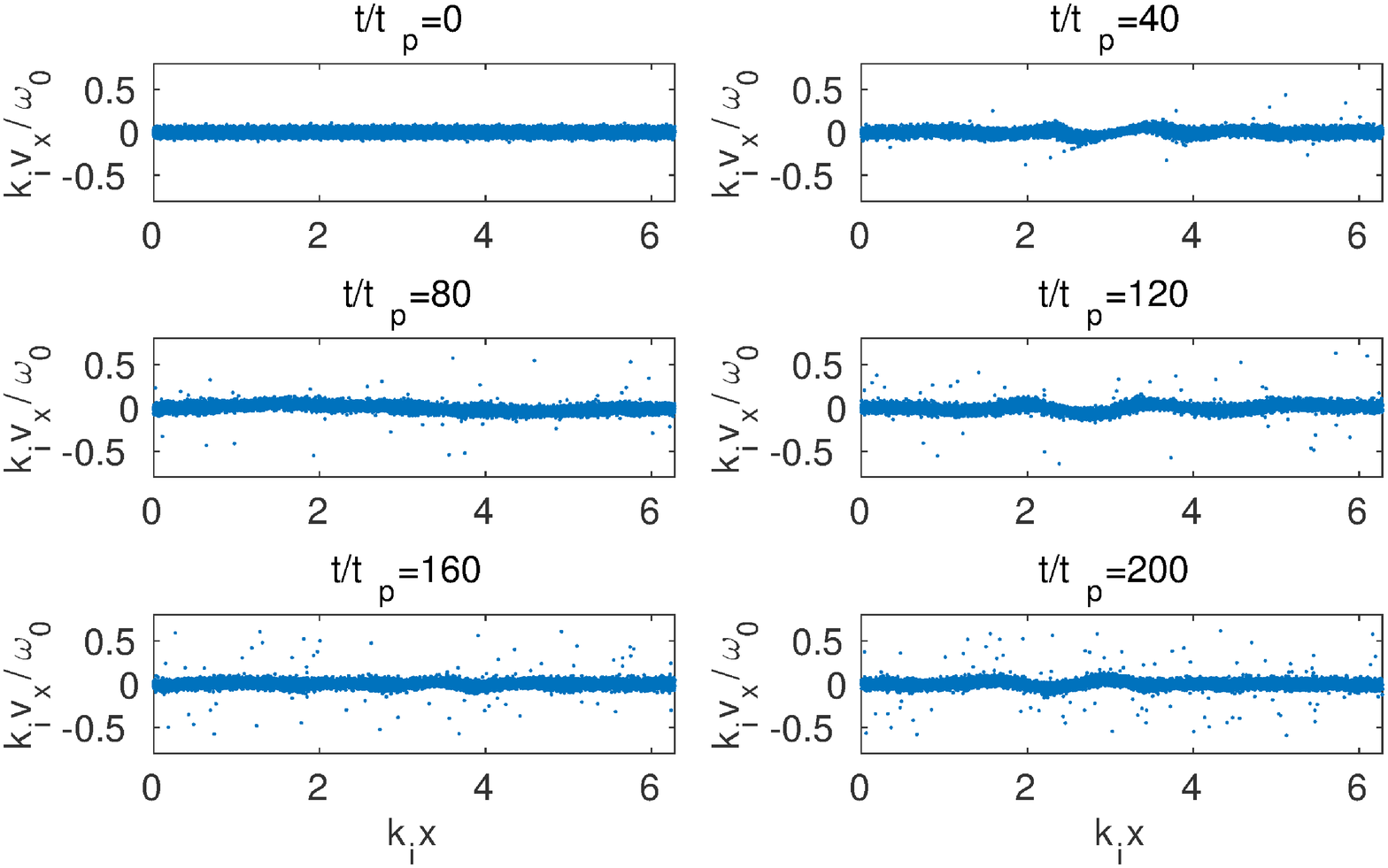}
\caption{Phase space plot for $v_{th}=0.03$ and $\alpha=0.05$ ($k_{max} < k_{wb}$).}
\label{fig:9}
\includegraphics[width=1.0\linewidth,height=0.5\linewidth]{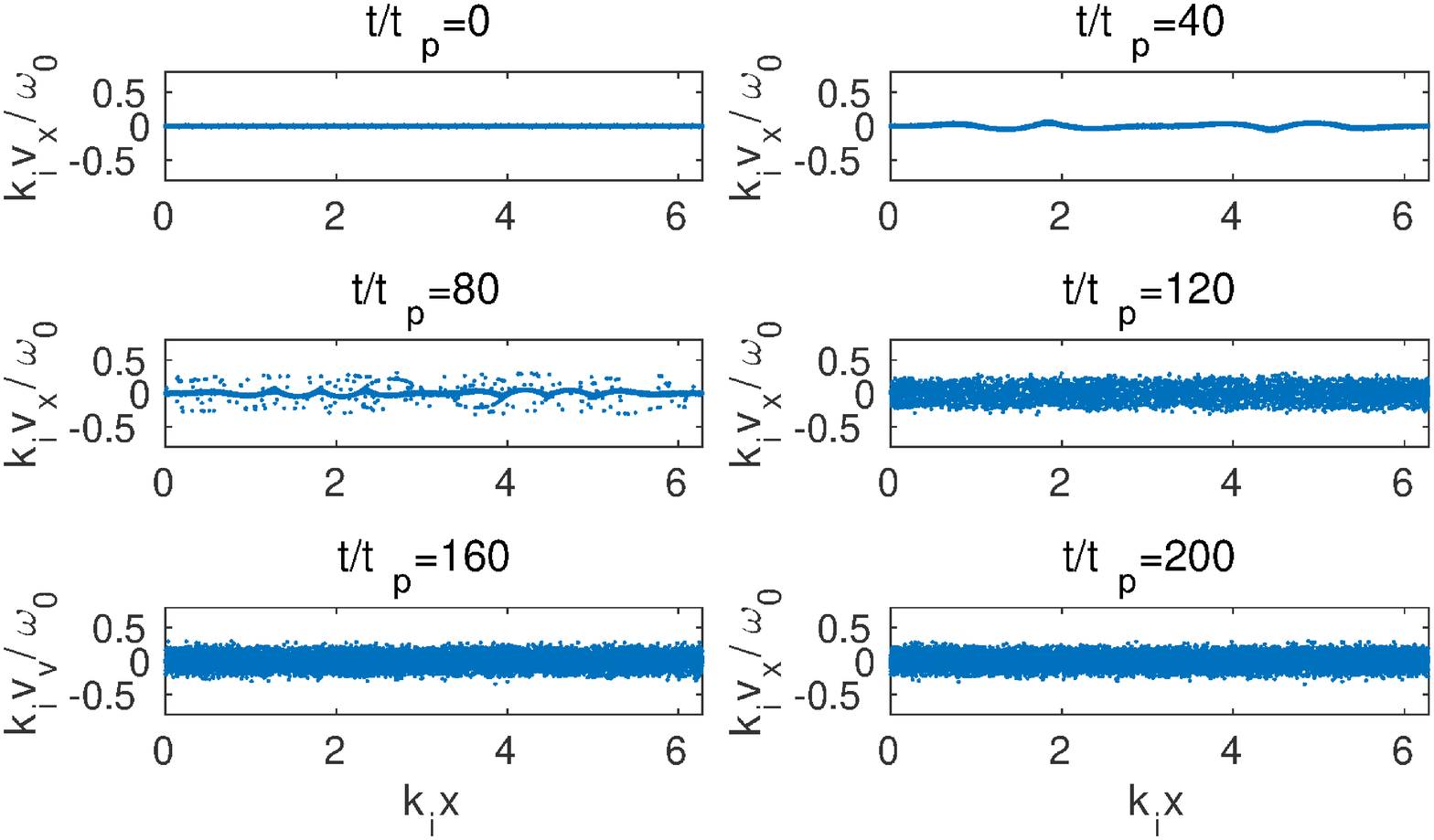}
\caption{Phase space plot for $v_{th}=0.005$ and $\alpha=0.05$ ($k_{max} > k_{wb}$).}
\label{fig:10}
\end{figure}

We now present snapshots of the electron phase space for two different values of $v_{th}$ and for a fixed value of $\alpha$. Fig.\ref{fig:9} and \ref{fig:10} respectively show the phase space plots for $v_{th} = 0.03$ ($k_{max} < k_{wb}$) and  $v_{th}=0.005$ ($k_{max} > k_{wb}$). In both two cases $\alpha$ is kept fixed at $0.05$. By comparing these two figures we observe, as expected, that for $v_{th}=0.005$ wave breaks as a large fraction of the wave energy is transferred to the particles as random kinetic energy (as a result phase space broadens), whereas for $v_{th}=0.03$ wave does not break as the phase space does not exhibit generation of significant number of energetic particles. 

\section{Summary}\label{summary}

In this paper we have studied the space time evolution of electron plasma oscillations in a warm inhomogeneous plasma. It is argued that for an inhomogeneous plasma, there exists a critical temperature beyond which wave breaking does not occur. Using an $1-D$ PIC simulation, this claim has been verified. It is shown that interplay between thermal pressure and background inhomogeneity restricts the wave number growth to a maximum value, $k_{max}=\sqrt{\alpha/\gamma}/\lambda_{D}$. This $k_{max}$ essentially governs the dynamics of the excited wave and governs whether it will break or walk-off. For cases where $k_{max} > k_{wb}$ wave breaks and transfers it's energy to particles as random thermal energy where as for $k_{max} < k_{wb}$ wave does not break ( walks-off ) even in the presence of inhomogeneity. It is further shown that in the presence of electron temperature, electron density never becomes singular even at the time of breaking.  The present work thus clearly brings out the distinction between wave breaking and walk-off. We specifically point to the phase space plots; although the peak density is finite in both cases i.e. whether the wave breaks or ``walks off'', the phase space plots give a completely different picture.

The results presented are of relevance to experiments where large amplitude plasma waves are excited, for e.g. laser - plasma interaction experiments. We note here that our work does not directly apply to plasma waves which are driven by an external driver. In case of externally driven plasma waves, for e.g. plasma waves driven by an intense laser pulse, the amplitude of the excited wave at the critical layer, not only depends on the amplitude of the driver but also on the background inhomogeneity scale length, whereas the wavelength depends on the inhomogeneity\cite{kruer88}. The electrons at the critical layer undergo forced oscillations in the presence of the driver. In our simulations, although we do not have an external driver ( electrons do not execute forced oscillations), the amplitude of the plasma wave, which is self-excited because of the choice of our initial conditions, depends on the amplitude and scale length of the background inhomogeneity and the wavelength depends on the inhomogeneity scale length. The conditions used are thus partly representative of the actual scenario in laser-plasma interactions, and leads to a basic understanding of how electron plasma oscillations respond to background ion inhomogeneity and thermal pressure of electrons. Simulations of plasma waves in an inhomogeneous plasma with an external driver are left for future studies.

\nocite{*}
\bibliography{paper2}
\bibliographystyle{unsrt}

\end{document}